\documentclass[aps,prc,letterpaper,11pt,twoside,tightenlines,nofootinbib,showpacs,preprint]{revtex4}
\usepackage{graphicx}
\usepackage[sort&compress]{natbib}
\usepackage{subfigure}
\usepackage{amsmath}
\usepackage{amsfonts}
\usepackage{cancel}

\newcommand{\Ima}{\textrm{Im}}

\newcommand{\mev}{\textrm{ MeV}}
\newcommand{\be}{\begin{equation}}
\newcommand{\ee}{\end{equation}}
\newcommand{\ba}{\begin{eqnarray}}
\newcommand{\ea}{\end{eqnarray}}

\begin{document}

\title{The small $K \pi$ component in the $K^*$ wave functions}

\author{C. W. Xiao $^1$, F. Aceti $^1$ and M. Bayar $^{1,\;2}$}

\affiliation{$^1$Departamento de F\'{\i}sica Te\'orica and IFIC, Centro Mixto Universidad \\de Valencia-CSIC, Institutos de Investigaci\'on de Paterna, Apartado 22085, 46071 Valencia, Spain\\
$^2$ Department of Physics, Kocaeli University, 41380 Izmit, Turkey}

\date{\today}

\begin{abstract}
We use a recently developed formalism which generalizes the Weinberg's compositeness condition to partial waves higher than $s$-wave in order to determine the probability of having a $K \pi$ component in the $K^*$ wave function. A fit is made to the $K \pi$ phase shifts in $p$-wave, from where the coupling of $K^*$ to $K \pi$ and the $K \pi$ loop function are determined. These ingredients allow us to determine that the $K^*$ is a genuine state, different to a  $K \pi$ component, in a proportion of about 80 \%.

\end{abstract}

%\pacs{13.75.Lb; 14.40.Lb; 21.45.-v}

\maketitle

\section{Introduction}

Understanding the nature and structure of hadronic particles is an important subject of hadron physics. In principle Quantum Chromodynamics (QCD) should give an answer to these questions. At high energies, because of the asymptotic freedom \cite{Gross:1973id,Politzer:1973fx,Politzer:1974fr}, QCD can be treated perturbatively, but at low energies, needed to  interpret the hadron spectrum, QCD is highly non perturbative and calculations become very difficult. Lattice QCD can provide an answer in the future, but so far the determination of the hadron spectrum is finding more problems than anticipated, in particular for particles which decay in several channels, which are the majority of them \cite{sasha,fu}.

Traditionally quark models have tried to find an approach to that problem \cite{Feynman:1964fk, isgur,capstick,Valcarce:2005em} and remarkable progress has been done from this perspective, but in order to understand the particle properties quoted in the Particle Data Group (PDG) \cite{pdg2012} it is also becoming clear that hadronic states are more complex than just three quarks for the baryons and $q \bar{q}$ for the mesons \cite{vijande}. One of the theories that has been remarkably successful dealing with hadron interaction and structure is chiral perturbation theory \cite{Weinberg:1978kz,Gasser:1983yg}. This theory is an effective field theoretical approach to QCD at low energies and the quark and gluon degrees of freedom are substituted with the baryons and mesons themselves. Yet, it soon became clear that chiral perturbation theory has a very limited energy range of convergence and improvements were made to construct non perturbative unitary extensions of the theory that allowed to deal with hadron interactions at much higher energies. These extensions are commonly known as the chiral unitary approach \cite{Kaiser:1995eg,Oller:1997ti,Oset:1997it,Oller:1998hw,Oller:1998zr, Oller:2000fj,Jido:2003cb,Guo:2006fu,Guo2006wp,GarciaRecio:2002td,GarciaRecio:2005hy,Hyodo:2002pk} (see \cite{Oller:2000ma} for a review). With this theory one can study the interaction between hadrons, and some times the interaction leads to poles in the scattering matrix which are interpreted  in terms of \textquoteleft\textquoteleft dynamically generated" or \textquoteleft \textquoteleft composite hadron" states, like the two $\Lambda(1405)$,  $\Lambda(1670)$, $N^*(1535)$, etc.

One of the questions that attracts attention is the issue of whether some resonances are \textquoteleft \textquoteleft composite" of other hadrons or \textquoteleft \textquoteleft genuine" states (other than hadron-hadron molecular states). One answer to this question was early given in the paper of Weinberg  \cite{Weinberg4} (see also \cite{Baru:2003qq, Hanhart:2010wh, Cleven:2011gp}), but it deals with particles bound in $s$-wave with a very small binding. The generalization to also $s$-waves but not necessarily so lightly bound and with many coupled channels was given in \cite{gamerjuan} for bound states and it was extended to deal with resonances in \cite{yamajuan}. A further generalization to higher partial waves was done in \cite{Acetifirst}. In this latter paper it was found that the $\rho$ meson had a $\pi\pi$ component in the wave function that amounted only to about 20 \%, which allows one to claim that the $\rho$ is basically a genuine state rather than a composite state of $\pi\pi$.

In the present work we want to extend the work done in \cite{Acetifirst} for the $\rho$ meson to the $K^*$. The $K^*$ particle was first reported by \cite{Wojcicki:1964zz} and also confirmed in \cite{Dauber:1967zza,Barash:1900zz}. It is always exhibited as a resonance in the $K \pi$ scattering \cite{Mercer:1971kn,Estabrooks:1977xe}, which is determined from experiments by the reactions $K^{\pm} p \to K^{\pm} \pi^+ n$, $K^{\pm} p \to K^{\pm} \pi^- \Delta^{++}$ and $K^+ p \to K^0 \pi^0 \Delta^{++}$. The $p$-wave $K \pi$ scattering was studied by N/D method in \cite{Oller:1998zr} using the chiral unitary approach and good agreement was found between theory and experiment. But there are no works focusing on the structure of the $K^*$ resonance from the point of view of its possible $K \pi$ compositeness or otherwise and this is the aim of our present work.

\section{Brief summarize of formalism}
\label{formal}

Following \cite{Acetifirst} one uses the set of coupled Schr\"{o}dinger equations
\begin{equation}
\begin{split}
\label{eq:LSe}
|\Psi\rangle &=|\Phi\rangle +\frac{1}{E-H_{0}}V|\Psi\rangle \\
&=|\Phi\rangle +\frac{1}{E-M_i - \frac{\vec{p}\;^2}{2 \mu_i}}V|\Psi\rangle\ ,
\end{split}
\end{equation}
where $H_{0}$ is the free Hamiltonian,  $\mu_{i}$ is the reduced mass of the system of total mass $M_{i}=m_{ai}+m_{bi}$, and
\begin{equation}
|\Psi\rangle=
\begin{Bmatrix}
|\Psi_{1}\rangle \\ |\Psi_{2}\rangle \\\vdots \\|\Psi_{N}\rangle
\end{Bmatrix}\ ,\ \ \ \ \ \ \
|\Phi\rangle=
\begin{Bmatrix}
|\Phi_{1}\rangle \\0\\\vdots \\0
\end{Bmatrix}\ ,
\end{equation}
where $|\Phi_1\rangle$ represents the only channel present at $t=- \infty$ taken as a plane wave. $V$ is the potential chosen as
\be 
\langle\vec{p}|V|\vec{p}\ '\rangle \equiv (2l+1)\ v\ \Theta(\Lambda-p)\Theta(\Lambda-p')|\vec{p}\,|^{l}|\vec{p}\ '|^{l}P_{l}(\cos\theta)\ ,\label{eq:v}
\ee
where $v$ is a $N\times N$ matrix with $N$ the number of channels, $\Lambda$ is a cutoff in the momentum space, and $P_{l}(\cos\theta)$ is the Legendre function. Note that the functions inherent to a $l$-wave character have been explicitly taken into account in $V$ and hence $v$ is considered as a constant in Eq. \eqref{eq:v}.

As shown in \cite{Acetifirst} one can see that the $T$-matrix can be written in the form of  Eq. \eqref{eq:v} and one finds
\be 
\langle\vec{p}|T|\vec{p}\ '\rangle \equiv (2l+1)\ t\ \Theta(\Lambda-p)\Theta(\Lambda-p')|\vec{p}\,|^{l}|\vec{p}\ '|^{l}P_{l}(\cos\theta)\ ,\label{eq:t}
\ee
with
\begin{equation}
\label{eq:scatmatrix}
t=(1-vG)^{-1}v\ ,
\end{equation}
where $G$ is the loop function for the two intermediate hadron states, which differs technically from the one more commonly used in the chiral unitary approach \cite{Oller:1998zr} in that it contains the factor $|\vec{p}\,|^{2l}$ in the integral, since this factor has been removed from $v$ (see Eq. \eqref{eq:v}). Hence
\begin{equation}
\label{eq:Gii}
G_{ii}=\int_{|\vec{p}\,|<\Lambda}d^{3}\vec{p}\;\frac{|\vec{p}\,|^{2l}}{E-M_i-\frac{\vec{p}\,^2}{2\mu_{i}}+i\epsilon}\ .
\end{equation}
Following again \cite{Acetifirst}, one finds that for a resonance or bound state, which is dynamically generated by the interaction, the sum rule
\begin{equation}
\label{eq:sumrule2}
-\sum_{i}g_{i}^{2}\left[\frac{dG_{i}}{dE}\right]_{E=E_p} = 1
\end{equation}
is fulfilled, with $E_p$ the position of the complex pole. However, if the state contains some genuine component outside the space of the $N$ wave functions of the coupled channels approach, Eq. \eqref{eq:sumrule2} is generalized to 
\begin{equation}
-\sum_{i}g_{i}^{2}\left[\frac{dG_{i}}{dE}\right]_{E=E_p} + |\langle\beta|\Psi\rangle|^2 = 1\ ,
\end{equation}
or equivalently to
\begin{equation}
\label{eq:sumrule}
-\sum_{i}g_{i}^{2}\left[\frac{dG_{i}}{dE}\right]_{E=E_p} = 1-Z\ ;\ \ \ \ \ \ \ Z=|\langle\beta|\Psi\rangle|^2\ ,
\end{equation}
where $|\beta\rangle$ is the genuine component of the state, and $g_{i}$ is the coupling, defined as
\begin{equation}
g_{i}g_{j}=\lim_{E\rightarrow E_p}(E-E_p)\ t_{ij}\ .
\end{equation}

\section{The $K\pi$ scattering in $p$-wave and the $K^*$ resonance}

Now we investigate the structure of the $K^*$ particle, which shows up as a resonance of $K \pi$, using the formalism discussed before with just one channel. In order to quantify the statement, we first start from the chiral unitary approach, and then we use a pure phenomenological method which is independent from any theoretical model to confirm our results.

\subsection{Chiral unitary model}

The $p$-wave $K \pi$ scattering is studied in \cite{Oller:1998zr} using the N/D method and only the tree level scattering potential. We take the potential from \cite{Oller:1998zr} but taking into account the definition of Eq. \eqref{eq:v}, thus removing the three-momentum factor,
\begin{equation}
v=-\frac{1}{2 f^2}\left(1+\frac{2G_V^2}{f^2}\frac{s}{M_{K^*}^{2}-s}\right)\ ,
\end{equation}
where $M_{K^*}$ is the bare $K^*$ mass, $f$ is the $\pi$ decay constant and $G_V$ the coupling to $K \pi$ in the formalism of \cite{Gasser:1983yg}, where $G_V \simeq f/\sqrt{2}$.

\begin{figure}
\centering
%\includegraphics[scale=0.6]{piKDelta0fit2.eps}
%\includegraphics[scale=0.6]{piKDelta0.eps}
%\caption{The fit results of the $K \pi$ scattering $p$-wave phase shift. Left: the fit of experiments data points; Right: the fit results. The data are taken from \cite{Mercer:1971kn,Estabrooks:1977xe,Oller:1998zr}.}
\subfigure[\ The fit of experimental data points.]{\label{fig:fig1a}\includegraphics[scale=0.6]{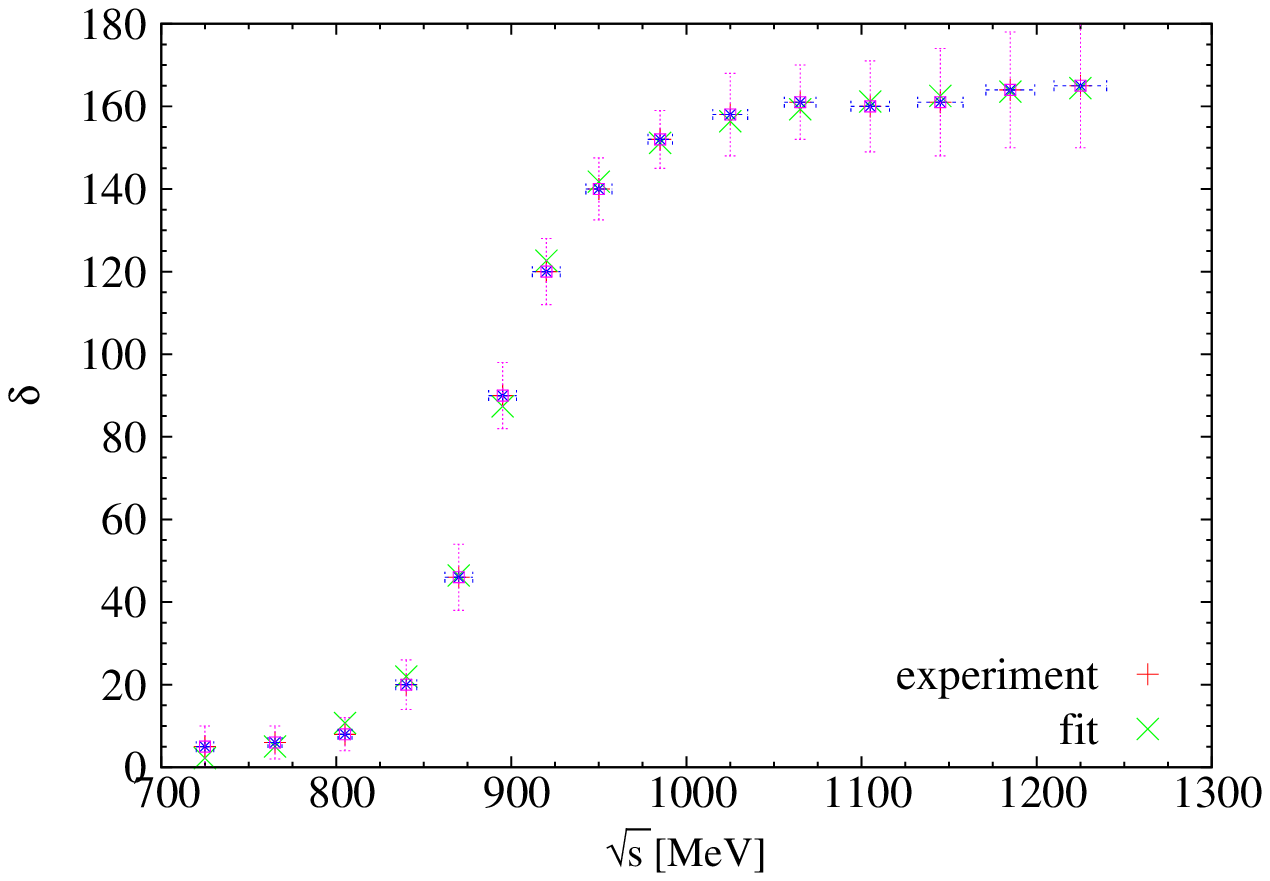}}
\subfigure[\ The fit results.]{\label{fig:fig1b}\includegraphics[scale=0.6]{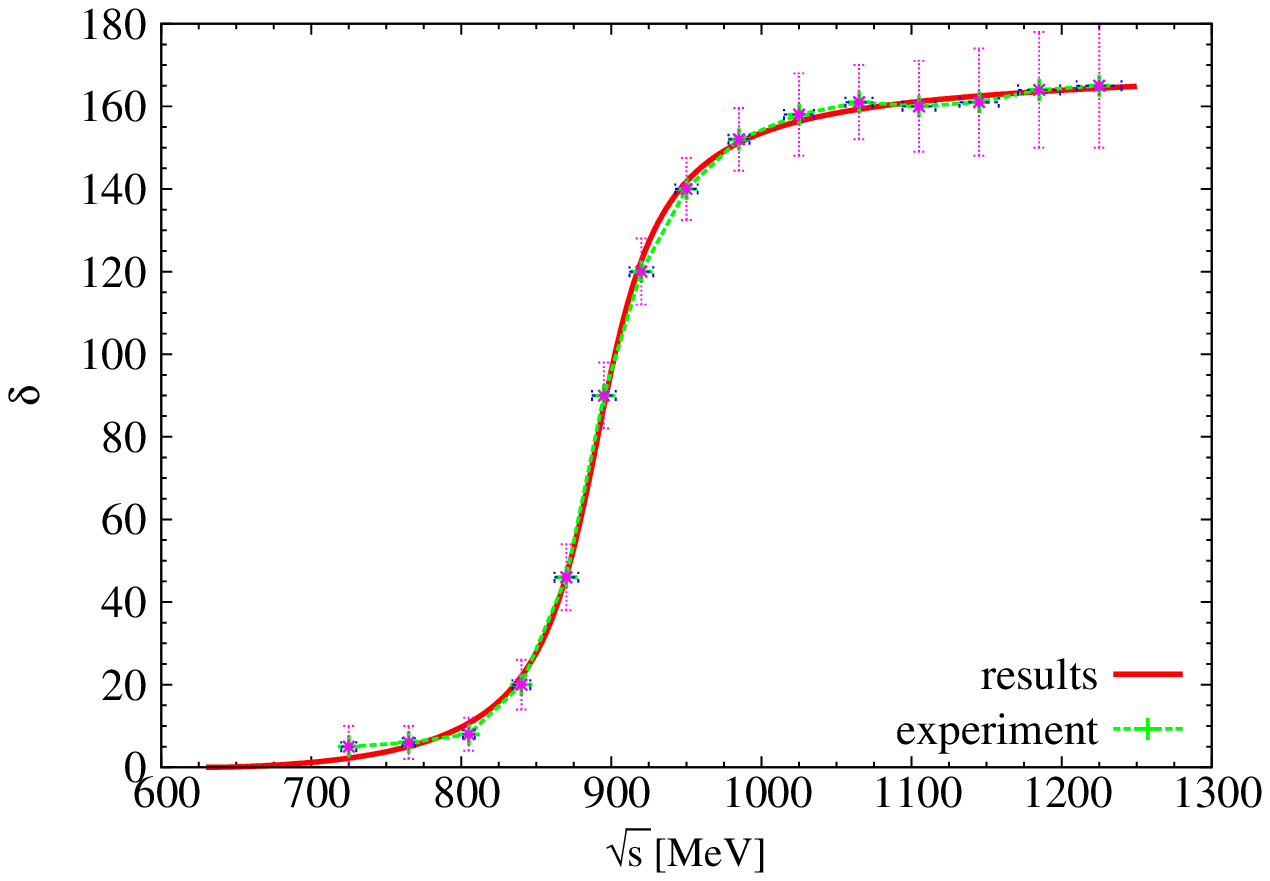}}
\caption{The fit results of the $K \pi$ scattering $p$-wave phase shift. The data are taken from \cite{Mercer:1971kn,Estabrooks:1977xe,Oller:1998zr}.}
\label{fig:fig1}
\end{figure}

\begin{figure}
\centering
\includegraphics[scale=0.6]{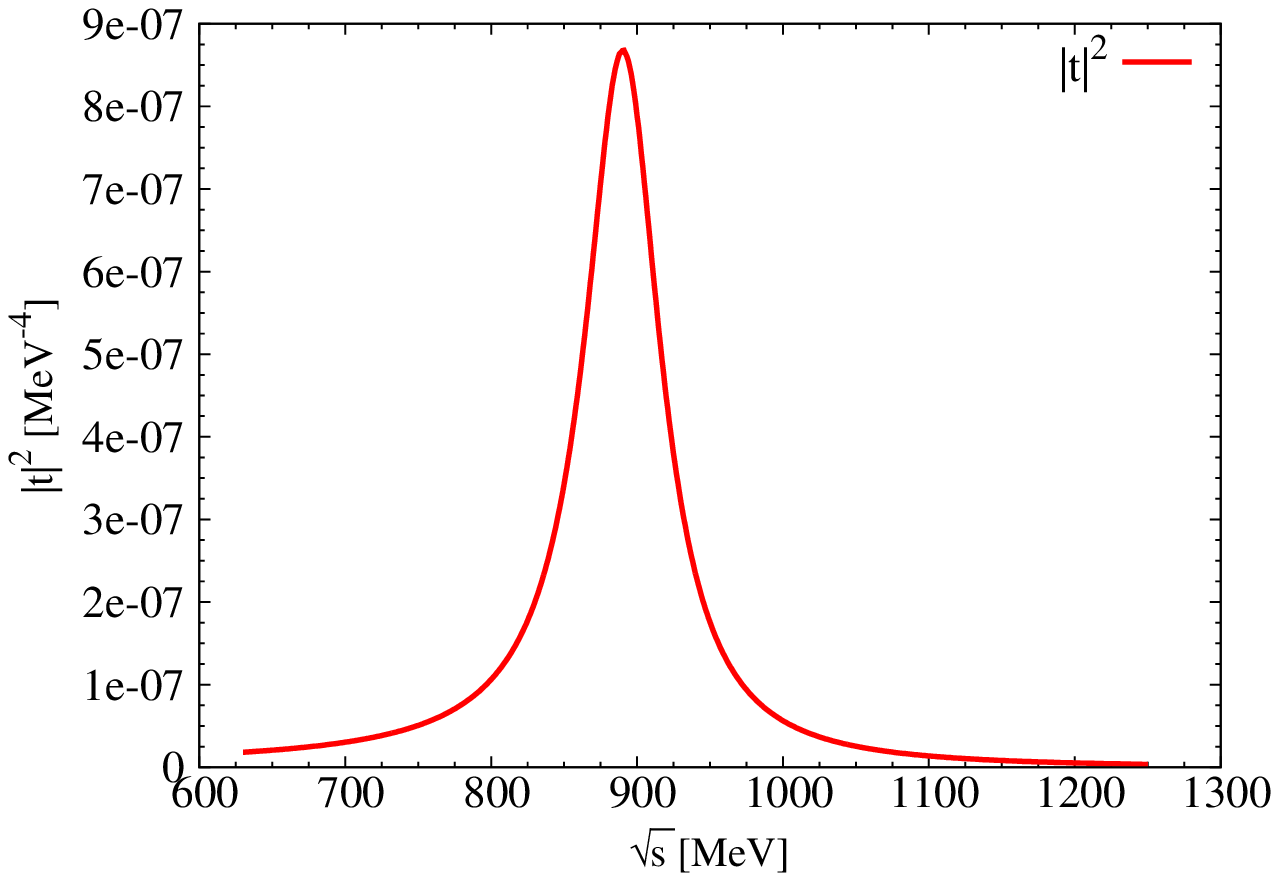}
\includegraphics[scale=0.6]{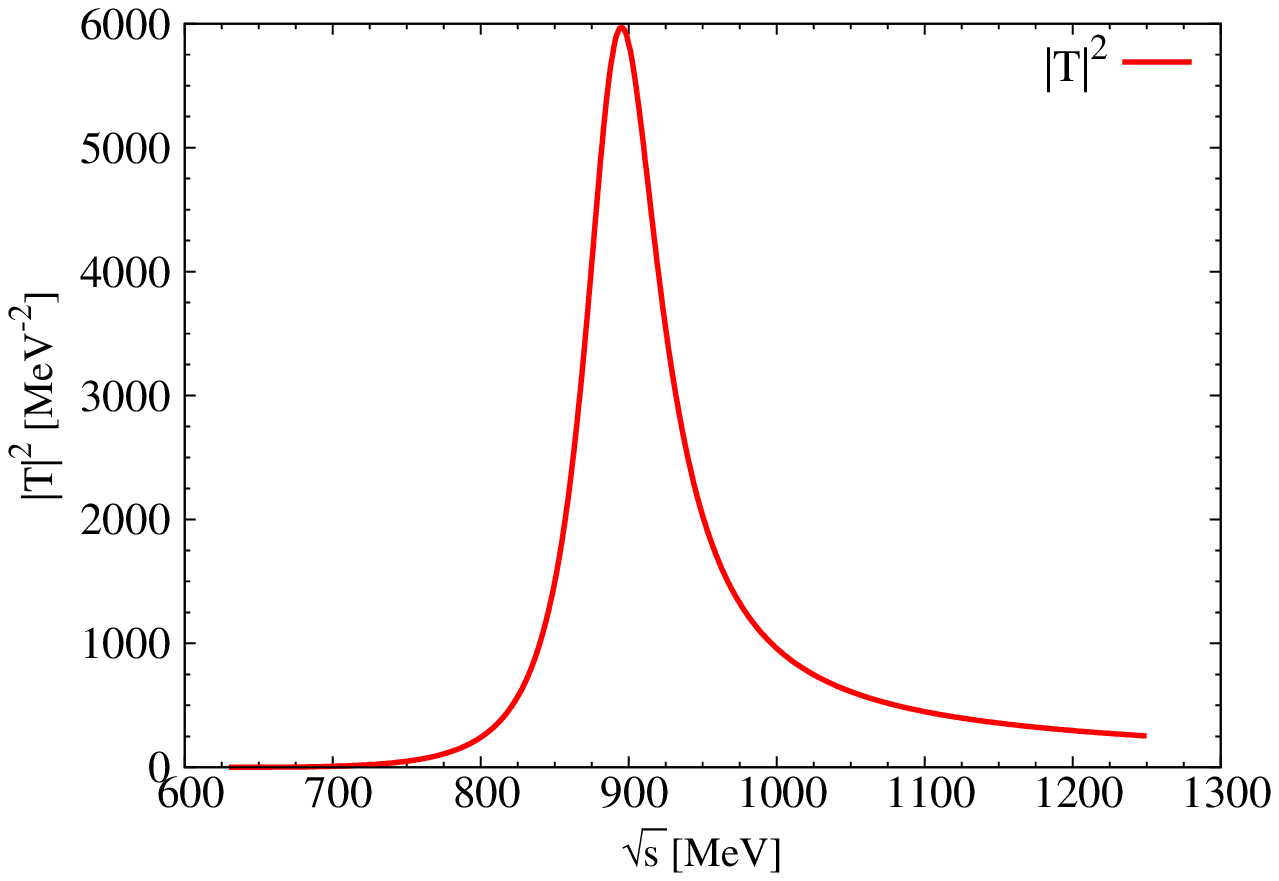}
\caption{Modulus squared of the $K \pi$ scattering amplitudes. Left: $|t|^2$; Right: $|T|^2$ with  $cos \theta = 1$.}
\label{fig:fig2}
\end{figure}

Here, we have explicitly separated the factor $|\vec{p}\,|^2$ in the potential $V$ to get $v$ which does not depend on the momentum. The formalism using this $v$ kernel requires, as shown in the Sec. \ref{formal}, that the $|\vec{p}\,|^2$ factor should be included in the loop function (see Eq. \eqref{eq:Gii} and the discussion in \cite{Acetifirst}). Thus, we can fit the data \cite{Mercer:1971kn,Estabrooks:1977xe,Oller:1998zr} by using Eq. \eqref{eq:scatmatrix} in one channel,
\begin{equation}
\label{eq:ampl_test1}
t=\frac{v}{1-vG}\ .
\end{equation}
As also done in \cite{Acetifirst}, we generalize the formulas of the former section to make them relativistic defining $g_{i}g_{j}$ as
\begin{equation}
\label{eq:gigj2}
g_{i}g_{j}=\lim_{s\rightarrow s_R}(s-s_R)\ t_{ij}\ ,
\end{equation}
with the loop function $G$ given by
\begin{equation}
\label{eq:G_test1}
G(s)=\int_{|\vec{q}\,|<q_{max}}{\frac{d^3 \vec{q}}{(2\pi)^3}}\frac{|\vec{q}\,|^2}{s-(\omega_{1}+\omega_{2})^2+i\epsilon}\left(\frac{\omega_{1}+\omega_{2}}{2\omega_{1}\omega_{2}}\right)\ ,
\end{equation}
where $\omega_i=\sqrt{m_i^2+\vec{q}\,^2},\ i=\pi,\ K$. The loop function of Eq. \eqref{eq:G_test1} is regularized by means of a cutoff $q_{max}$. The $p$-wave $K \pi$ phase shift is then given by the formula \cite{Acetifirst}
\begin{equation}
p^2\,t=\frac{-8\pi\sqrt{s}}{p\cot{\delta(p)}-i\,p}\ ,
\end{equation}
with $p$ the three-momentum in the center of mass reference frame, which is given by
\begin{equation}
\label{eq:pcm}
p=\frac{\lambda^{1/2}(s,m^2_1,m^2_2)}{2\sqrt{s}}
\end{equation}
where $\lambda(s,m^2_1,m^2_2)=[s-(m_1+m_2)^2]\cdot[s-(m_1-m_2)^2]$. Then we carry a $\chi^2$ fit to the data \cite{Mercer:1971kn,Estabrooks:1977xe,Oller:1998zr} using the parameters $f,\ G_V,\ M_{K^*},\ q_{max}$, with $f,\ G_V$ constrained not to differ much from standard values. For the best fit of the data, we find the values of these free parameters:
\begin{equation}
\begin{split}
&f=86.22\mev\ , \\&
G_V=53.81\mev\ , \\&
M_{K^*}=995.76\mev\ , \\&
q_{max}=724.698\mev\ .
\end{split}
\end{equation}
The best fit results are shown in Fig. \ref{fig:fig1}. In Fig. \ref{fig:fig1a} we show the best fit to data points for the $K\pi$ phase shift which are taken from the experimental data \cite{Mercer:1971kn,Estabrooks:1977xe,Oller:1998zr}. Fig. \ref{fig:fig1b} shows the results of the fit using the determined parameters for a continuum of energies.
Using the fit parameters, we also get the results for the modulus squared of the scattering amplitudes $|t|^2$ and $|T|^2$, which are shown in Fig. \ref{fig:fig2}.
From Fig. \ref{fig:fig2}, we can see a clear peak in the modulus squared of the amplitudes which corresponds to a resonant structure, with a mass about $890\mev$ and a width about $50\mev$. In order to apply the sum rule to the case of a resonance, we should extrapolate the amplitude to the complex plane and look for the complex pole $s_0$ in the second Riemann sheet. This is done by changing $G$ to $G^{II}$ in Eq. (\ref{eq:ampl_test1}) to get the complex amplitude in the second Riemann sheet, $t^{II}$. We proceed as follows:

$G^{II}(s)$ is the analytic continuation to the complex plane of the loop function \cite{Oller:1997ti} in $p$-wave, given by
\begin{equation}
G^{II}(s)=G^{I}(s)+i\frac{p^3}{4\pi\sqrt{s}}\ ,\ \ \ \ \ \ \ \ \ \ \ \Ima (p)> 0\ ,
\end{equation} 
where $G^{I}$ and $G^{II}$ are the loop functions in the first and second Riemann sheet, $G^{I}$ is given by Eq. (\ref{eq:G_test1}), and $p$ is the complex momentum in the center of mass reference frame, given by Eq. \eqref{eq:pcm}. In the second Riemann sheet, we find the pole of the resonance by solving the equation
\begin{equation}
1-vG=0\ .
\end{equation}
Thus we are now able to determine the coupling $\tilde{g}_{\rho}$ as the residue in the pole of the amplitude using Eq. \eqref{eq:gigj2}, as
\begin{equation}
\label{eq:gcouple}
g^2=\lim_{s\to s_{0}}(s-s_{0})t^{II}\\ . 
\end{equation}
Finally, we can use the sum rule of Eq. \eqref{eq:sumrule} for the single $K \pi$ channel (generalized to the relativistic case) in order to evaluate the contribution of this channel to the wave function of the resonance,
\begin{equation}
\label{eq:sr}
-g^2 \left[\frac{dG^{II}(s)}{ds}\right]_{s=s_{0}}=1-Z\ ,
\end{equation}
where $Z$ represents the probability that the $K^*$ is not a $K \pi$ molecule but something else.

Using the determined parameters from the best fit to the data, we find the pole in the second Riemann sheet, which corresponds to the $K^*$ particle,
\begin{equation}
\sqrt{s_0}=(891.0+i\ 31.3)\mev\ ,
\end{equation}
which is consistent with the results of Fig. \ref{fig:fig2} and the PDG group \cite{pdg2012}, while the value of the coupling is
\begin{equation}
g=g_{K \pi}=(7.19+i\ 0.67)\ .
\end{equation}
Then, we get
\begin{equation}
\begin{split}
&1-Z=(0.122+i\ 0.193)\ ,\\ &|1-Z|=0.229\ ,
\end{split}
\end{equation}
which indicates that the amount of $K \pi$ in the wave function is small. One can conclude that the $K^*$ is largely a genuine state other than a $K\pi$ composite molecule.

\subsection{Phenomenological analysis}

As done in \cite{Acetifirst}, we also use  a pure phenomenological analysis to confirm our results only with  experimental data. The case for a $p$-wave resonance is different from the one for $s$-wave, where the coupling $g$ can be obtained from experiments and $\frac{dG}{dE}$ (or $\frac{dG}{ds}$) is a convergent magnitude, even when $q_{max}\to\infty$.

The phenomenological scattering amplitude in a relativistic form for $p$-wave can be written as 
\begin{equation}
\label{eq:Tphe}
\tilde{t}=\frac{\tilde{g}_{ex}^2}{s-m_{K^*}^{2} +i\;m_{K^*}\Gamma_{on}\left(\frac{p}{p_{on}}\right)^{3}}\ ,
\end{equation}
where $p$ is the three-momentum of the $K \pi$ system in the center of mass reference frame, which is given by Eq. \eqref{eq:pcm} for real $\sqrt{s}$, and $p_{on}$ is the same quantity for $\sqrt{s} = m_{K^*}$,
\begin{equation}
p_{on}=p(\sqrt{s}=m_{K^*})\ ,
\end{equation}
and the coupling is related to the width through the equation
\begin{equation}
\tilde{g}_{ex}^2=\frac{8\pi m_{K^*}^2\Gamma_{on}}{p_{on}^3}\ .
\end{equation}
Besides, the values of the mass $m_{K^*}$ and width $\Gamma_{on}$ of the $K^*$ are given by experiment.

As done in the former subsection, to get the pole and the coupling, we also need to extrapolate the amplitude to the complex plane and search for the pole $s_0$ in the second Riemann sheet. We obtain $\tilde{t}$ in the second Riemann sheet from Eq. (\ref{eq:Tphe}) by taking $s$ complex, $s=a+i\ b$, and $p\rightarrow -p$ in the width term. Thus, we can look for the pole in the second Riemann sheet and then use Eq. \eqref{eq:gcouple} to evaluate the coupling. We obtain
\begin{equation}
\begin{split}
&\sqrt{s_0}=(892.0+i\ 22.4)\mev\ , \\
&\tilde{g}_{K \pi}=(6.08+i\ 0.50)\ ,
\end{split}
\end{equation}
which are consistent with those obtained before and closer to the value of experiment \cite{pdg2012} for the width.

For the $p$-wave, the $G$ function in Eq. \eqref{eq:G_test1} or Eq. \eqref{eq:Gii} is not convergent and $\frac{dG}{ds}$ is also logarithmically divergent \cite{Acetifirst, guo}. Therefore, when doing the $1-Z$ calculation, one does not know which value of the cutoff $q_{max}$ should be used to regularize the $G$ function. Hence, as done in \cite{Acetifirst}, we can use natural values of the cutoff and test if the results are stable or not for a certain range of $q_{max}$.

In Table \ref{tab:vert}, we show the results of our study of the strength $|1-Z|$ obtained for the $K^*$ by changing the cutoff $q_{max}$ around a certain reasonable range. As we can see, the results are stable and similar to those obtained in the former subsection. Particularly, for $|1-Z|$ we get the same conclusion as before, which means that, since $|1-Z|$ is a small number, the $K^*$ is not a $K \pi$ composite state.
\begin{table}[ tp ]
\begin{tabular}{c|c|c}
\hline %
$q_{max}\ [\textrm{MeV}]$ &  $1-Z$ & $|1-Z|$\\\toprule %
$724.7$ & $0.082-i0.137$ & $0.160$ \\
$700.0$ & $0.077-i0.138$ & $0.158$ \\
$800.0$ & $0.095-i0.134$ & $0.165$ \\
$900.0$ & $0.111-i0.131$ & $0.172$ \\
$1000.0$ & $0.124-i0.128$ & $0.179$ \\
$1100.0$ & $0.136-i0.126$ & $0.186$ \\
$1200.0$ & $0.147-i0.124$ & $0.192$ \\\hline
\end{tabular}
\caption{Values of $1-Z$ for different cutoffs $q_{max}$.}
\label{tab:vert}\centering %
\end{table}

\section{Conclusions}

In the present work, we show the results of our investigation of the $K^*$ structure. We use a recently developed method for studying the particle structure by the generalized Weinberg's compositeness condition, which extends the results of Weinberg for $s$-wave to other partial waves and bound states or resonances. Using this formalism, we first calculate the $K \pi$ coupling with the chiral unitary theory, by means of the tree level chiral potential, and then we make a fit to the experimental data of the phase shifts to determine the free parameters in this model. With the best fit to the experimental data we get the pole of the $K^*$ resonance in the second Riemann Sheet, $(891.0 + i\;31.3)\mev$, which is consistent with the PDG data. With the coupling of $K \pi$, we find that the probability of the $K \pi$ component, $|1-Z|$, is a small value, only $0.229$ (about 1/5), which means that the $K^*$ is not a $K \pi$ molecule but something else. Next, we also use a phenomenological method to confirm the former results. The pole in the second Riemann sheet is $(892.0 + i\;22.4)\mev$ and $|1-Z|$ has values around $0.158 \sim 0.192$, which are in agreement with the results of the theoretical model analysis within uncertainties, thus, leading to the same conclusion.

\section*{Acknowledgements}

We thank E. Oset for suggesting this problem and valuable help, and J.A. Oller for providing the data and useful discussion. One of us, C. W. X., thanks M.J. Vicente-Vacas for useful discussion and information, and also C. Hanhart, F.K. Guo for helpful suggestion in the Hadron Physics Summer School 2012 (HPSS2012), Germany. 
This work is partly supported by DGICYT contract number FIS2011-28853-C02-01, and the Generalitat Valenciana in the program Prometeo, 2009/090. We acknowledge the support of the European Community-Research Infrastructure Integrating Activity Study of Strongly Interacting Matter (acronym HadronPhysics3, Grant Agreement n. 283286) under the Seventh Framework Programme of the EU.
One of us, M. Bayar acknowledges support through the Scientific and Technical Research Council (TUBITAK) BIDEP-2219 grant.


\begin{thebibliography}{99}

%\cite{Gross:1973id}
\bibitem{Gross:1973id} 
  D.~J.~Gross and F.~Wilczek,
  %``Ultraviolet Behavior of Nonabelian Gauge Theories,''
  Phys.\ Rev.\ Lett.\  {\bf 30}, 1343 (1973).
  %%CITATION = PRLTA,30,1343;%%

 %\cite{Politzer:1973fx}
\bibitem{Politzer:1973fx} 
  H.~D.~Politzer,
  %``Reliable Perturbative Results for Strong Interactions?,''
  Phys.\ Rev.\ Lett.\  {\bf 30}, 1346 (1973).
  %%CITATION = PRLTA,30,1346;%%

%\cite{Politzer:1974fr}
\bibitem{Politzer:1974fr} 
  H.~D.~Politzer,
  %``Asymptotic Freedom: An Approach to Strong Interactions,''
  Phys.\ Rept.\  {\bf 14}, 129 (1974).
  %%CITATION = PRPLC,14,129;%%
 
%\cite{Lang:2012sv}
\bibitem{sasha} 
  C.~B.~Lang, L.~Leskovec, D.~Mohler and S.~Prelovsek,
  %``K pi scattering for isospin 1/2 and 3/2 in lattice QCD,''
  Phys.\ Rev.\ D {\bf 86}, 054508 (2012)
  [arXiv:1207.3204 [hep-lat]].
  %%CITATION = ARXIV:1207.3204;%%

%\cite{Fu:2012tj}
\bibitem{fu}
  Z.~Fu and K.~Fu,
 %``Lattice QCD study on $K^\ast(892)$ meson decay width,''
 arXiv:1209.0350 [hep-lat].
 %%CITATION = ARXIV:1209.0350;%%   
  
%\cite{Feynman:1964fk}
\bibitem{Feynman:1964fk} 
  R.~P.~Feynman, M.~Gell-Mann and G.~Zweig,
  %``Group U(6) x U(6) generated by current components,''
  Phys.\ Rev.\ Lett.\  {\bf 13}, 678 (1964).
  %%CITATION = PRLTA,13,678;%%
  
%\cite{Isgur:1978xj}
\bibitem{isgur}
  N.~Isgur and G.~Karl,
 %``P Wave Baryons in the Quark Model,''
 Phys.\ Rev.\ D {\bf 18}, 4187 (1978).
 %%CITATION = PHRVA,D18,4187;%%

%\cite{Capstick:2000qj}
\bibitem{capstick}
  S.~Capstick and W.~Roberts,
 %``Quark models of baryon masses and decays,''
 Prog.\ Part.\ Nucl.\ Phys.\  {\bf 45}, S241 (2000)
 [nucl-th/0008028].
 %%CITATION = NUCL-TH/0008028;%%

%\cite{Valcarce:2005em}
\bibitem{Valcarce:2005em}
  A.~Valcarce, H.~Garcilazo, F.~Fernandez and P.~Gonzalez,
 %``Quark-model study of few-baryon systems,''
 Rept.\ Prog.\ Phys.\  {\bf 68}, 965 (2005)
 [hep-ph/0502173].
 %%CITATION = HEP-PH/0502173;%%   
  
%\cite{Beringer:1900zz}
\bibitem{pdg2012} 
  J.~Beringer {\it et al.}  [Particle Data Group Collaboration],
  %``Review of Particle Physics (RPP),''
  Phys.\ Rev.\ D {\bf 86}, 010001 (2012).
  %%CITATION = PHRVA,D86,010001;%%  
  
%\cite{Vijande:2003ki}
\bibitem{vijande}
  J.~Vijande, F.~Fernandez, A.~Valcarce and B.~Silvestre-Brac,
 %``Tetraquarks in a chiral constituent quark model,''
 Eur.\ Phys.\ J.\ A {\bf 19}, 383 (2004)
 [hep-ph/0310007].
 %%CITATION = HEP-PH/0310007;%%

%\cite{Weinberg:1978kz}
\bibitem{Weinberg:1978kz}
  S.~Weinberg,
 %``Phenomenological Lagrangians,''
 Physica A {\bf 96}, 327 (1979).
 %%CITATION = PHYSA,A96,327;%%the

%\cite{Gasser:1983yg}
\bibitem{Gasser:1983yg}
  J.~Gasser and H.~Leutwyler,
 %``Chiral Perturbation Theory to One Loop,''
 Annals Phys.\  {\bf 158}, 142 (1984).
 %%CITATION = APNYA,158,142;%%   
 
%\cite{Kaiser:1995eg}
\bibitem{Kaiser:1995eg} 
  N.~Kaiser, P.~B.~Siegel and W.~Weise,
  %``Chiral dynamics and the low-energy kaon - nucleon interaction,''
  Nucl.\ Phys.\ A {\bf 594}, 325 (1995)
  [nucl-th/9505043].
  %%CITATION = NUCL-TH/9505043;%%
  
%\cite{Oller:1997ti}
\bibitem{Oller:1997ti} 
  J.~A.~Oller and E.~Oset,
  %``Chiral symmetry amplitudes in the S wave isoscalar and isovector channels and the sigma, f0(980), a0(980) scalar mesons,''
  Nucl.\ Phys.\ A {\bf 620}, 438 (1997)
  [Erratum-ibid.\ A {\bf 652}, 407 (1999)]
  [hep-ph/9702314].
  %%CITATION = HEP-PH/9702314;%%
    
%\cite{Oset:1997it}
\bibitem{Oset:1997it} 
  E.~Oset and A.~Ramos,
  %``Nonperturbative chiral approach to s wave anti-K N interactions,''
  Nucl.\ Phys.\ A {\bf 635}, 99 (1998)
  [nucl-th/9711022].
  %%CITATION = NUCL-TH/9711022;%%
  
%\cite{Oller:1998hw}
\bibitem{Oller:1998hw} 
  J.~A.~Oller, E.~Oset and J.~R.~Pelaez,
  %``Meson meson interaction in a nonperturbative chiral approach,''
  Phys.\ Rev.\ D {\bf 59}, 074001 (1999)
  [Erratum-ibid.\ D {\bf 60}, 099906 (1999)]
  [Erratum-ibid.\ D {\bf 75}, 099903 (2007)]
  [hep-ph/9804209].
  %%CITATION = HEP-PH/9804209;%%
  
%\cite{Oller:1998zr}
\bibitem{Oller:1998zr} 
  J.~A.~Oller and E.~Oset,
  %``N/D description of two meson amplitudes and chiral symmetry,''
  Phys.\ Rev.\ D {\bf 60}, 074023 (1999)
  [hep-ph/9809337].
  %%CITATION = HEP-PH/9809337;%%
  
%\cite{Oller:2000fj}
\bibitem{Oller:2000fj} 
  J.~A.~Oller and U.~G.~Meissner,
  %``Chiral dynamics in the presence of bound states: Kaon nucleon interactions revisited,''
  Phys.\ Lett.\ B {\bf 500}, 263 (2001)
  [hep-ph/0011146].
  %%CITATION = HEP-PH/0011146;%%
    
%\cite{Jido:2003cb}
\bibitem{Jido:2003cb} 
  D.~Jido, J.~A.~Oller, E.~Oset, A.~Ramos and U.~G.~Meissner,
  %``Chiral dynamics of the two Lambda(1405) states,''
  Nucl.\ Phys.\ A {\bf 725}, 181 (2003)
  [nucl-th/0303062].
  %%CITATION = NUCL-TH/0303062;%%    

%\cite{Guo:2006fu}
\bibitem{Guo:2006fu} 
  F.~-K.~Guo, P.~-N.~Shen, H.~-C.~Chiang, R.~-G.~Ping and B.~-S.~Zou,
  %``Dynamically generated 0+ heavy mesons in a heavy chiral unitary approach,''
  Phys.\ Lett.\ B {\bf 641}, 278 (2006)
  [hep-ph/0603072].
  %%CITATION = HEP-PH/0603072;%%
  
%\cite{Guo:2005wp}
\bibitem{Guo2006wp}
  F.~K.~Guo, R.~G.~Ping, P.~N.~Shen, H.~C.~Chiang and B.~S.~Zou,
  %``S wave K pi scattering and effects of kappa in J/psi ---> anti-K*0 (892) K+
  %pi-,''
  Nucl.\ Phys.\  A {\bf 773}, 78 (2006).
%  [arXiv:hep-ph/0509050].
  %%CITATION = NUPHA,A773,78;%%  
  
%\cite{GarciaRecio:2002td}
\bibitem{GarciaRecio:2002td}
 C.~Garcia-Recio, J.~Nieves, E.~Ruiz Arriola and M.~J.~Vicente Vacas,
 %``S = -1 meson baryon unitarized coupled channel chiral perturbation theory and the S(01) Lambda(1405) and Lambda(1670) resonances,''
 Phys.\ Rev.\ D {\bf 67}, 076009 (2003)
 [hep-ph/0210311].
 %%CITATION = HEP-PH/0210311;%%

%\cite{GarciaRecio:2005hy}
\bibitem{GarciaRecio:2005hy} 
 C.~Garcia-Recio, J.~Nieves and L.~L.~Salcedo,
 %``SU(6) extension of the Weinberg-Tomozawa meson-baryon Lagrangian,''
 Phys.\ Rev.\ D {\bf 74}, 034025 (2006)
 [hep-ph/0505233].
 %%CITATION = HEP-PH/0505233;%%

%\cite{Hyodo:2002pk}
\bibitem{Hyodo:2002pk} 
 T.~Hyodo, S.~I.~Nam, D.~Jido and A.~Hosaka,
 %``Flavor SU(3) breaking effects in the chiral unitary model for meson baryon scatterings,''
 Phys.\ Rev.\ C {\bf 68}, 018201 (2003)
 [nucl-th/0212026].
 %%CITATION = NUCL-TH/0212026;%%
 
%\cite{Oller:2000ma}
\bibitem{Oller:2000ma} 
  J.~A.~Oller, E.~Oset and A.~Ramos,
  %``Chiral unitary approach to meson meson and meson - baryon interactions and nuclear applications,''
  Prog.\ Part.\ Nucl.\ Phys.\  {\bf 45}, 157 (2000)
  [hep-ph/0002193].
  %%CITATION = HEP-PH/0002193;%% 
  
%\cite{Weinberg:1965zz}
\bibitem{Weinberg4} 
  S.~Weinberg,
  %``Evidence That the Deuteron Is Not an Elementary Particle,''
  Phys.\ Rev.\  {\bf 137}, B672 (1965).
  %%CITATION = PHRVA,137,B672;%%

%\cite{Baru:2003qq}
\bibitem{Baru:2003qq} 
  V.~Baru, J.~Haidenbauer, C.~Hanhart, Y.~.Kalashnikova and A.~E.~Kudryavtsev,
  %``Evidence that the a(0)(980) and f(0)(980) are not elementary particles,''
  Phys.\ Lett.\ B {\bf 586}, 53 (2004)
  [hep-ph/0308129].
  %%CITATION = HEP-PH/0308129;%%

%\cite{Hanhart:2010wh}
\bibitem{Hanhart:2010wh} 
  C.~Hanhart, Y.~.S.~Kalashnikova and A.~V.~Nefediev,
  %``Lineshapes for composite particles with unstable constituents,''
  Phys.\ Rev.\ D {\bf 81}, 094028 (2010)
  [arXiv:1002.4097 [hep-ph]].
  %%CITATION = ARXIV:1002.4097;%%
  
%\cite{Cleven:2011gp}
\bibitem{Cleven:2011gp} 
  M.~Cleven, F.~-K.~Guo, C.~Hanhart and U.~-G.~Meissner,
  %``Bound state nature of the exotic $Z_b$ states,''
  Eur.\ Phys.\ J.\ A {\bf 47}, 120 (2011)
  [arXiv:1107.0254 [hep-ph]].
  %%CITATION = ARXIV:1107.0254;%%  
  
%\cite{Gamermann:2009uq}
\bibitem{gamerjuan}
 D.~Gamermann, J.~Nieves, E.~Oset, E.~Ruiz Arriola,
 %``Couplings in coupled channels versus wave functions: application to the X(3872) resonance,''
 Phys.\ Rev.\  {\bf D81}, 014029 (2010).
% [arXiv:0911.4407 [hep-ph]].

%\cite{YamagataSekihara:2010pj}
\bibitem{yamajuan}
 J.~Yamagata-Sekihara, J.~Nieves, E.~Oset,
 %``Couplings in coupled channels versus wave functions in the case of resonances: application to the two $\Lambda(1405)$ states,''
 Phys.\ Rev.\  {\bf D83}, 014003 (2011).
% [arXiv:1007.3923 [hep-ph]].

%\cite{Aceti:2012dd}
\bibitem{Acetifirst} 
  F.~Aceti and E.~Oset,
  %``Wave functions of composite hadron states and relationship to couplings of scattering amplitudes for general partial waves,''
  Phys.\ Rev.\ D {\bf 86}, 014012 (2012)
  [arXiv:1202.4607 [hep-ph]].
  %%CITATION = ARXIV:1202.4607;%%


%\cite{Wojcicki:1964zz}
\bibitem{Wojcicki:1964zz} 
  S.~G.~Wojcicki,
  %``K- p --> Kbar0 pi- p Reaction from 1.0 to 1.7 GeV/c,''
  Phys.\ Rev.\  {\bf 135}, B484 (1964).
  %%CITATION = PHRVA,135,B484;%%

%\cite{Dauber:1967zza}
\bibitem{Dauber:1967zza} 
  P.~M.~Dauber, P.~E.~Schlein, W.~E.~Slater and H.~K.~Ticho,
  %``Reaction K- p --> K- p pi+ pi- at 2.0 GeV/c,''
  Phys.\ Rev.\  {\bf 153}, 1403 (1967).
  %%CITATION = PHRVA,153,1403;%%
  
%\cite{Barash:1900zz}
\bibitem{Barash:1900zz} 
  N.~Barash, L.~Kirsch, D.~Miller and T.~H.~Tan,
  %``Annihilations of Antiprotons at Rest in Hydrogen. VI. Kaonic Final States,''
  Phys.\ Rev.\  {\bf 156}, 1399 (1967).
  %%CITATION = PHRVA,156,1399;%%
    
%\cite{Mercer:1971kn}
\bibitem{Mercer:1971kn} 
  R.~Mercer, P.~Antich, A.~Callahan, C.~Y.~Chien, B.~Cox, R.~Carson, D.~Denegri and L.~Ettlinger {\it et al.},
  %``K pi scattering phase shifts determined from the reactions k+ p ---> k+ pi- delta++ and k+ p ---> k0 pi0 delta++,''
  Nucl.\ Phys.\ B {\bf 32}, 381 (1971).
  %%CITATION = NUPHA,B32,381;%%

%\cite{Estabrooks:1977xe}
\bibitem{Estabrooks:1977xe} 
  P.~Estabrooks, R.~K.~Carnegie, A.~D.~Martin, W.~M.~Dunwoodie, T.~A.~Lasinski and D.~W.~G.~S.~Leith,
  %``Study of K pi Scattering Using the Reactions K+- p ---> K+- pi+ n and K+- p ---> K+- pi- Delta++ at 13-GeV/c,''
  Nucl.\ Phys.\ B {\bf 133}, 490 (1978).
  %%CITATION = NUPHA,B133,490;%%

\bibitem{guo}
  F. K. Guo, lectures in the Hadron Physics Summer School 2012 (HPSS2012), Germany.



  
\end{thebibliography}
\end{document}